\documentclass[%
reprint,
superscriptaddress,
groupedaddress,
frontmatterverbose, 
nofootinbib,
 amsmath,amssymb,
 aps,
]{revtex4-1}

\usepackage{graphicx}
\usepackage{dcolumn}
\usepackage{bm}
\usepackage{hyperref}
\usepackage[mathlines]{lineno}

\usepackage[latin1]{inputenc}
\usepackage{amsfonts}

\begin{document}


\title{Solutions of the Einstein field equations for a bounded and finite discontinuous source, and its generalization: Metric matching conditions and jumping effects}


\author{Ramon Lapiedra}
 \email{ramon.lapiedra@uv.es}
\affiliation{%
Departament d'Astronomia i Astrof\'{\i}sica, Universitat de
Val\`encia, 46100 Burjassot, Val\`encia, Spain
}%
\affiliation{%
Observatori Astron\`omic, Universitat de
Val\`encia, E-46980 Paterna, Val\`encia, Spain.
}%
\author{Juan Antonio Morales-Lladosa}%
 \email{antonio.morales@uv.es}
\affiliation{%
Departament d'Astronomia i Astrof\'{\i}sica, Universitat de
Val\`encia, 46100 Burjassot, Val\`encia, Spain
}%
\affiliation{%
Observatori Astron\`omic, Universitat de
Val\`encia, E-46980 Paterna, Val\`encia, Spain.
}%

\date{\today}

\begin{abstract}
We consider the metrics of the General Relativity, whose energy-momentum tensor has a bounded support where it is continuous except for a finite step across the corresponding boundary surface. As a consequence, the first derivative of the metric across this boundary could perhaps present a finite step too. However, we can assume that the metric is ${\cal C}^1$ class everywhere. In such a case, although the partial second derivatives of the metric exhibit finite (no Dirac $\delta$ functions) discontinuities, the Dirac $\delta$ functions will still appear in the conservation equation of the energy-momentum tensor. As a consequence, strictly speaking, the corresponding metric solutions of the Einstein field equations can only exist in the sense of distributions. Then, we assume that the metric considered is ${\cal C}^1$ class everywhere and is a solution of the Einstein field equations in this sense.
We explore the consequences of these two assumptions, and in doing so we derive the general conditions that constrain the jumps in the second partial derivatives across the boundary. The example of the Oppenheimer-Snyder metric is considered and some new results are obtained on it. Then, the formalism developed in this exploration is applied to a different situation, i.e., to a given generalization of the Einstein field equations for the case where the partial second derivatives of the metric exist but are not symmetric.
\end{abstract}


\maketitle



\section{Introduction: general considerations}
\label{intro}
In a well-known formalization of the general relativity Lichnerowicz \cite{Lichnerowicz} assumes the two following postulates: (1) The space-time manifold is ${\cal C}^2$ class, that is, in any two overlapping charts, the corresponding coordinate transformation functions and their first and second derivatives are continuous.(2) In admissible coordinates, the metric is required to be ${\cal C}^1$ class, and its first derivatives are required to be piecewise ${\cal C}^2$ class.%
\footnote{Here, the required piecewise ${\cal C}^2$ class character of the first derivatives of the metric has the following specific sense: These first derivatives are simply continuous everywhere instead of being piecewise continuous. This is a consequence of having assumed that the metric is ${\cal C}^1$  class.\label{trossos}}
This ${\cal C}^2$ character means in particular that, out of some discontinuity surface, the second and third derivatives of the metric exist and are continuous such that, out of this surface, the Einstein field equations and the Bianchi identities, respectively, can be written out.
By definition, admissible coordinates are such that the second derivatives of the above coordinate transformation functions are piecewise ${\cal C}^2$ class; that is, their third and fourth derivatives exist but could be discontinuous through some 3-surfaces. This kind of coordinate definition is consistent with the %
above postulate (2) for the metric. Notice that, out of the boundary surface, the postulate allows, in particular, for the existence of continuous second and third derivatives of the metric. Then, this continuous character remains preserved, irrespective of the coordinates used, provided that we use admissible coordinates in the sense just defined, since then the resultant Jacobian matrix will be piecewise ${\cal C}^3$ class. See also \cite{Lake}. 

There is a dubious justification for this postulated ${\cal C}^1$ character of the metric in the current literature \cite{Synge}: In the Newtonian limit,  the ten gravitational potentials reduce to the Newtonian potential satisfying the Poisson equation, whose physical solution for a finite and bounded source is well known to be ${\cal C}^1$ class. Later, however,  Lichnerowicz \cite{LichnerowiczB} required only the continuity of the metric when studying shock gravitational waves.

In any case, if the metric is continuous but it is not ${\cal C}^1$ class everywhere, the second derivatives of the metric will be somewhere Dirac $\delta$ {\it functions}, and then this metric could be a solution of the Einstein field equations (EFEs) only in the sense of distributions, or otherwise said,  a {\it weak} solution of these equations.

All this makes it interesting to clarify in which specific circumstances we could state the ${\cal C}^1$ postulate for the metric in an admissible coordinate system. One of the aims of the present paper is to address such a clarification, but before entering perfunctorily into it in this Introduction, let us briefly comment some references:

In Ref. \cite{Mansouri}, a given hypersurface $\Sigma$ is considered where the energy-momentum tensor has a part that behaves like some Dirac $\delta$ function: a ``thin shell" or ``surface layer". Then, in the same reference, the corresponding {\it weak} metric solutions of the EFEs, continuous across the $\Sigma$ hypersurface, are considered. In the particular case of a ``boundary surface", characterized by a finite step in the energy density, and thus a case where this Dirac $\delta$ part vanishes, the metric solutions lead to the Darmois-Israel matching conditions, that is, the continuity of the intrinsic metric and the extrinsic curvature across the boundary surface \cite{Darmois,Israel}. Some precedents on the subject can be found in \cite{Gemelli}. See more recently \cite{Senovilla}, where some previous results on the subject of the matching conditions and the field equations in the sense of distributions are generalized.

It is easy to see that requiring these Darmois-Israel conditions in Gauss coordinates leads to the ${\cal C}^1$ character of the metric in these coordinates (see, for instance, \cite{Lake}). This does not mean that the above postulate (2) of Lichnerowicz on this metric character will be necessarily satisfied, since this postulate requires in particular that the coordinates used be admissible ones (see again \cite{Lake}). However, the reciprocal is true: if this postulate 2) of Lichnerowicz was satisfied, the metric would be ${\cal C}^1$ class in Gauss coordinates, and these Gauss coordinates would be admissible too (see p. 61 of \cite{Lichnerowicz}).

Let us go back to our clarification attempt  mentioned above: Under what specific circumstances could we postulate that a metric solution of EFEs is ${\cal C}^1$ class? As stated in the Abstract, the metrics considered in the present paper will be the ones with a bounded support for the energy-momentum tensor $T_{\alpha\beta}$, this tensor being everywhere continuous except for a finite step across its boundary surface. Despite the presence of the support boundary surface, in the present paper we assume the metric to be ${\cal C}^1$ class across this boundary surface, and we explore the consequences of such an assumption. Notice then that, because of the EFEs, the second derivatives of such a metric across the boundary surface will not present Dirac $\delta$  functions, but only finite jumps (Sec. \ref{sec:2}). However, some Dirac $\delta$ functions will still appear in the equation expressing the conservation of the energy-momentum tensor because of the finite discontinuities of this tensor. Thus, in all, the assumed kind of metric could only be the solution of the EFEs in the sense of distributions, that is, a {\it weak} solution of these equations. Hence, in line with Ref. \cite{Mansouri}, we require our metrics to satisfy EFEs in this sense. As we will see, this will be compatible with our ${\cal C}^1$ class metric assumption and will give us some specific information about the discontinuities across the boundary surface of the metric second derivatives in relation to the discontinuities of the energy-momentum tensor and its first derivatives (Secs. \ref{sec:3} and \ref{sec:4}). In all, one of our results here is that, for the kind of energy-momentum tensor that we are considering, there is always a large family of metric solutions of the EFEs that satisfy the ${\cal C}^1$ class metric postulate in some coordinates that could be admissible ones. This result is compatible with the above-mentioned result (see, for instance, \cite{Berrocoso}) of the theory of the gravitational potential, that the standard solution of the Poisson equation for a suitable source  is ${\cal C}^1$ class. On the other hand, it can be compared with the result we commented above: For a boundary surface (diferent of the above defined ``surface layer" or its equivalent ``thin shell''), the matching Darmois-Israel conditions are satisfied. This comparison needs to be made having in mind the relation between these matching conditions and the Lichnerowicz ones (see \cite{Lake}).

For the sake of completeness, we refer here to the items considered in Secs. \ref{sec:5} and \ref{sec:6}. In Sec. \ref{sec:5}, we consider the example of the Oppenheimer-Snyder (OS) metric \cite{Oppenheimer} in the coordinates of Szekeres \cite{Szekeres} where this metric exhibits explicitly its ${\cal C}^1$ class character, and we prove that these coordinates are admissible ones. Further, in Sec. \ref{sec:6}, leaving the question approached up to now, i.e., the one related to the stated Lichnerowicz conditions, we apply the formalism developed in this approach to a different problem concerning the generalization of the EFEs for the case where the second partial derivatives of the metric exist but do not commute \cite{Kuntz}, that is, the case in which the Schwartz theorem is no longer valid. Then, we find the relatively simple equations to which the jumps of these second partial derivatives obey and comment on its possible interest. Finally, in Sec. \ref{sec:7} we summarize our findings. In the Appendix we make explicit some calculations in relation to the OS metric in Szekeres coordinates, and we prove that the corresponding Jacobian is everywhere nonvanishing, and further that the first derivatives of the metric are piecewise ${\cal C}^2$ class, two results that are not present in Ref. \cite{Szekeres}. In all, the essential result obtained, jointly in Sec. \ref{sec:5} and Appendix, is that the OS metric in Szekeres coordinates satisfies the Postulate (2) of Lichnerowicz.

We use signature +2. Greek indices take values from $0$ to $3$ and Latin indices from $1$ to $3$. The gravitational constant and the speed of light are taken equal to $1$. The sign conventions adopted when defining the curvature and Ricci tensors are the ones used in Ref. \cite{MTW-Gravitation}.


\section{The EFEs and its discontinuous part}
\label{sec:2}

Let it be the EFEs:
\begin{equation}
R_{\alpha\beta}= 8\pi (T_{\alpha\beta}-\frac{1}{2}T_{\gamma}^{\gamma} g_{\alpha\beta})\equiv  8\pi S_{\alpha\beta}, \label{EFE}
\end{equation}
where with the suitable global sign the Ricci tensor becomes
\begin{eqnarray}\label{EFEexplicit}
R_{\alpha\beta} & = &  \frac{1}{2}g^{\mu\kappa}(\partial^2_{\alpha\kappa}g_{\mu\beta} -  \partial^2_{\mu\kappa}g_{\alpha\beta} - \partial^2_{\alpha\beta}g_{\mu\kappa} + \partial^2_{\mu\beta}g_{\kappa\alpha}) \nonumber \\ & & 
+ g^{\lambda\mu}g_{\eta\sigma}(\Gamma_{\beta\lambda}^\eta \Gamma_{\alpha\nu}^\sigma - \Gamma_{\nu\lambda}^\eta \Gamma_{\alpha\beta}^\sigma), 
\end{eqnarray}
where $\partial^2_{\alpha\beta}g_{\mu\nu} \equiv \partial_\alpha \partial_\beta g_{\mu\nu}$.  
As stated in the Introduction, we will consider a bounded energy-momentum tensor everywhere continuous except in the boundary surface, where it can present finite discontinuities, i.e. steps. Despite this, we will assume that the metric is ${\cal C}^1$ class and explore the consequences of this assumption. Then, in the EFE we will have discontinuous finite jump functions for the second metric derivatives, that will be present only in the part of the Ricci tensor, $R_{\alpha\beta}$, containing these second derivatives of the metric. We will write $R^D_{\alpha\beta}$ for this discontinuous part. Then, in accordance with (\ref{EFEexplicit}), we will have
\begin{equation}
R^D_{\alpha\beta} =\frac{1}{2}g^{\mu\kappa}(\partial^2_{\alpha\kappa}g_{\mu\beta} - \partial^2_{\mu\kappa}g_{\alpha\beta} -  \partial^2_{\alpha\beta}g_{\mu\kappa} + \partial^2_{\mu\beta}g_{\kappa\alpha}). \label{principal}
\end{equation}

Thus, across the boundary surface, the EFEs imply that the  $R^D_{\alpha\beta}$ discontinuity, must be equal to $8\pi$ times the discontinuity of the $S_{\alpha\beta}$ tensor. Thus, we write:
\begin{equation}
[R^D_{\alpha\beta}]= 8\pi[S_{\alpha\beta}].  \label{principal-Eq}
\end{equation}
More precisely, if $\Phi(x^\alpha)=0$ is the equation of the boundary surface, we define $[R^D_{\alpha\beta}]\equiv R^D_{\alpha\beta}|_{\Phi \to 0^{+}}-R^D_{\alpha\beta}|_{\Phi \to 0^{-}}$ and similarly for $[S_{\alpha\beta}]$.


\section{The energy-momentum conservation in the distributional sense}
\label{sec:3}

In the preceding sections we have assumed that the metric is ${\cal C}^1$ class across the boundary surface of the energy-momentum tensor. Consequently, the second derivatives of the metric present finite discontinuities across it to deal with the corresponding finite discontinuities of this tensor. Then, Dirac $\delta -{\it funcions}$ will appear in the third derivatives of the metric. This means that the divergence of the Einstein tensor, $G_{\alpha\beta}\equiv R_{\alpha\beta}-\frac{1}{2}R g_{\alpha\beta}$, only vanishes identically in the sense of distributions. Now, let us assume this kind of vanishing and consider the corresponding vanishing of the divergence of the energy-momentum tensor. We will have
\begin{equation}
\nabla_\alpha T^\alpha_\beta =0. \label{T-divergence-vanishing}
\end{equation}
Here, $T^{\alpha}_{\beta}$ presents a finite discontinuity across the boundary surface. Then, Dirac $\delta$ funcions will appear in the ordinary first derivatives of this tensor across this boundary, implying that (\ref{T-divergence-vanishing}) is only true in the sense of distributions.

Let us denote by $\nabla_\alpha T^\alpha_\beta (\delta)$ the part of $\nabla_\alpha T^\alpha_\beta $ containing the $\delta$ and only the $\delta$ terms. Obviously, because of (\ref{T-divergence-vanishing}), this part vanishes. On the other hand, since the metric is ${\cal C}^1$ class and $T_{\alpha\beta}$ only presents finite discontinuities, we will have in an obvious notation $\nabla_\alpha T^\alpha_\beta (\delta)=\partial_\alpha T^\alpha_\beta(\delta)$. Thus,
\begin{equation}
\partial_\alpha T^\alpha_\beta(\delta) =0. \label{T-delta-vanishing}
\end{equation}

Then, because of the finite discontinuity of $T^\alpha_\beta$ in its boundary, we can write
\begin{equation}
T^\alpha_\beta=T^\alpha_\beta(\Phi\leq 0)H_L(\Phi)+T^\alpha_\beta(\Phi>0)H_R(\Phi), \label{T-expression}
\end{equation}
where $H_L$ and $H_R$ are the corresponding Heaviside functions, that is,
\begin{equation}
H_L(\Phi \le 0)=1, H_L(\Phi>0)=0, \label{HL}
\end{equation}
and
\begin{equation}
H_R(\Phi \le 0)=0, H_R(\Phi>0)=1. \label{HR}
\end{equation}
Then, we have
\begin{eqnarray}
\partial_\alpha  T^\alpha_\beta & = & \partial_\alpha  T^\alpha_\beta (\Phi \le 0)H_L(\Phi)+\partial_\alpha  T^\alpha_\beta (\Phi>0)H_R(\Phi) \nonumber \\ 
& & + [T^\alpha_\beta]\delta(\Phi)\partial_\alpha \Phi, \label{ Tderivative}
\end{eqnarray}
where we have taken into account that $dH_L/d\Phi= -\delta(\Phi)$ and $dH_R/d\Phi= \delta(\Phi)$.

Hence, (\ref{T-delta-vanishing}) becomes
\begin{equation}
[T^\alpha_\beta]\partial_\alpha \Phi \delta(\Phi)=0, \label{T-step}
\end{equation}
that is,
\begin{equation}
[T^\alpha_\beta]n_\alpha =0,
 \label{without-delta}
\end{equation}
with $n_\alpha$ the unit vector $n_\alpha \equiv \frac{\partial_\alpha \Phi}{|\partial \Phi|} $, assumed to be spacelike, i.e., $g^{\alpha\beta}n_\alpha n_\beta=$1. The meaning of the notation $[T^\alpha_\beta]$, and other similar expressions henceforth, is the same as the one displayed in (\ref{principal-Eq}).


\section{The Hadamard discontinuities of the second derivatives of the metric and the Lichnerowicz postulate}
\label{sec:4}

In \cite{Hadamard}, Hadamard proves the following well-known result: let it be a function, $f(x^\alpha)$, continuous everywhere, whose first partial derivative, $\partial_\alpha f$, is finite and discontinuous across the boundary surface $\Phi(x^\alpha)=0$. Then, it is easy to see that
\begin{equation}
[\partial_\alpha f]=\kappa n_\alpha, \label{hadamard}
\end{equation}
where $\kappa$ is a given function of $x^\alpha$ defined on the boundary surface $\Phi=0$.

In our case, we have the functions $\partial_\lambda g_{\alpha\beta}$ that we have assumed to be continuous everywhere, while the second partial derivatives of the metric become piecewise continuous, that is, these second partial derivatives will have in general finite discontinuities across the boundary surface $\Phi=0$. Then, because of this Hadamard result, we will have
\begin{equation}
[\partial^2_{\lambda\mu} g_{\alpha\beta}]=\kappa_{\mu\alpha\beta} n_\lambda, \label{jumps}
\end{equation}
with $\kappa_{\mu\alpha\beta}$ some given functions of $x^\alpha$ defined on $\Phi=0$, symmetric in the $(\alpha,\beta)$ indices. But, assuming the Schwartz theorem about the mixed partial derivatives, $[\partial^2_{\lambda\mu} g_{\alpha\beta}]$ is, in particular, symmetric in both indices $\lambda$, $\mu$. Hence, (\ref{jumps}) becomes
\begin{equation}
[\partial^2_{\lambda\mu} g_{\alpha\beta}]=\kappa_{\alpha\beta} n_\lambda n_\mu, \label{jumps-B}
\end{equation}
where $\kappa_{\alpha\beta}$ stand for suitable functions of $x^\alpha$ defined on $\Phi=0$, symmetric in the $(\alpha,\beta)$ indices.

Then, substituting (\ref{jumps-B}) in (\ref{principal-Eq}), taking into account (\ref {principal}), we obtain
\begin{equation}
\kappa_{\lambda\mu}-\kappa_{\nu\mu} n^\nu n_\lambda- \kappa_{\nu\lambda}n^\nu n_\mu+\kappa n_\lambda n_\mu=-16\pi [S_{\lambda\mu}], \label{EFE-kappa}
\end{equation}
where $\kappa\equiv g^{\mu\nu}\kappa_{\mu\nu}$.

Now, the first question to consider is to check if Eqs. (\ref{EFE-kappa}) and (\ref{without-delta}) are compatible: we are going to see that the answer is positive. To begin with, in an obvious notation, (\ref{without-delta}) can be written as $[S^\alpha_\beta]n_\alpha -\frac{1}{2}[S]n_\beta =0$. Consequently, it should be $[G^\alpha_\beta]n_\alpha=0$, with $G^\alpha_\beta$ the Einstein tensor, that is, it should be
\begin{equation}
[R^\alpha_\beta]n_\alpha -\frac{1}{2}[R]n_\beta=0. \label{identity}
\end{equation}

But, having in mind the left-hand side of (\ref{EFE-kappa}), it is straightforward to verify that (\ref{identity}) is actually an identity. This completes the proof of the claimed compatibility. Notice by the way, that the algebraic Eq. (\ref{EFE-kappa}) actually allows for solutions. Thus, for instance, $\kappa_{\lambda\mu}=-16\pi[S_{\lambda\mu}]$, is a particular one. Another particular solution is considered in the next section.

In all, in the present case of a finite energy-momentum tensor with bounded support, allowing for a jump across the corresponding bounding surface, we have assumed the existence of local ${\cal C}^1$ class solutions of the EFEs. Then, we have been able to separate, from the corresponding general integration of these equations, the particular calculation of the second metric derivatives jumps on the boundary surface as linear functions of the corresponding steps in the energy-momentum tensor, in accordance with the algebraic relation (\ref{EFE-kappa}) (notice that although in general, in the EFEs, the second derivatives of the metric depend in particular on the first derivatives, it is not the case for the corresponding jumps).

This result becomes consistent with the above assumption, that is, with the assumed existence of ${\cal C}^1$ metric solutions. Actually, Eq. (\ref{EFE-kappa}) comes from Eq. (\ref{principal-Eq}), both equations being defined across $\Phi = 0$. The remaining EFEs across  $\Phi = 0$ are 
$(R_{\alpha\beta}) = 8 \pi (S_{\alpha\beta})$ with the notation 
$(R_{\alpha\beta}) = R_{\alpha\beta}|_{\Phi \to 0^{+}} +  R_{\alpha\beta}|_{\Phi \to 0^{-}}$
and similarly for $(S_{\alpha\beta})$. However, these equations do not involve the $[\partial^2_{\lambda\mu} g_{\alpha\beta}]$ jumps but, using the just defined notation, they only involves the counter parts $(\partial^2_{\lambda\mu} g_{\alpha\beta})$, that are independent of $[\partial^2_{\lambda\mu} g_{\alpha\beta}]$. In all, the complete EFEs across $\Phi = 0$ do not require any new conditions that could enter in contradiction with the above required ones, that is, with Eqs. (\ref{EFE-kappa}) and (\ref{without-delta}) plus the definition (\ref{jumps-B}). Then, since Eqs. (\ref{EFE-kappa}) and (\ref{without-delta}) are mutually compatible (actually, (\ref{without-delta}) is identically satisfied if (\ref{EFE-kappa}) is satisfied) and (\ref{EFE-kappa}) allows actually for solutions, Eq. (\ref{EFE-kappa}) becomes a necessary condition for having, in our case,  ${\cal C}^1$ class metric solutions. 

But, whether or not the coordinates in which this ${\cal C}^1$ character takes place are admissible coordinates will have to be tested  aside for every space-time considered. Similarly, the ${\cal C}^2$ character of the first derivatives of the metric [remember postulate (2))] should be tested too.
See the particular case of the OS metric in Szekeres coordinates \cite{Szekeres} at the end of the next section and in the Appendix, where the verification of  postulate (2) is finelly completed.


\section{Example of the Oppenheimer-Snyder metric}
\label{sec:5}

Let us consider the well-known solution of the EFEs, the OS metric \cite{Oppenheimer}. In \cite{Szekeres}, coordinate systems are derived where this metric is globally ${\cal C}^1$ class, but it is not shown to verify postulate (2). In this present section, we will apply to this particular case the previous general results obtained. But, beforehand, let us obtain a suitable family of explicit solutions of the algebraic Eq. (\ref{EFE-kappa}). In order to do this, we will use the following five symmetric tensors, $[T_{\alpha\beta}]$, $n_\alpha n_\beta$, $g_{\alpha\beta}(\Phi=0)$, $n_\alpha v_\beta +n_\beta v_\alpha$ and $v_\alpha v_\beta$, all them defined over the boundary surface $\Phi=0$, where $v_\alpha$ is a unit four-vector orthogonal to $n_\alpha$. Then, we will write for such solutions
\begin{eqnarray}
\kappa_{\alpha\beta} & = & X[T_{\alpha\beta}]+Yn_\alpha n_\beta+Z g_{\alpha\beta}(\Phi=0)
\nonumber \\ & & + P(n_\alpha v_\beta + n_\beta v_\alpha) + Q v_\alpha v_\beta, \label{general solution}
\end{eqnarray}
where $X$, $Y$, $Z$, $P$ and $Q$, are five functions, defined on $\Phi=0$, to be determined by the Eq. (\ref{EFE-kappa}). Substituting (\ref{general solution}) in (\ref{EFE-kappa}), we obtain:
\begin{eqnarray}
X[T_{\lambda\mu}] & + & Zg_{\lambda\mu}(\Phi=0)  +  Qv_\lambda v_\mu \nonumber \\ & + & (X[T_\alpha^\alpha] + 2Z + \epsilon Q) n_\lambda n_\mu 
 = -16\pi [S_{\lambda\mu}], \label{reducedEFE}
\end{eqnarray}
with $\epsilon \equiv g^{\alpha\beta}v_\alpha v_\beta =\pm1$, where we see that the $Y$ and $P$ functions do not appear. In other words, the two functions are arbitrary, expressing a remaining freedom in the use of the chosen coordinates. Furthermore, by contracting Eq. (\ref{reducedEFE}) with $n^\lambda$ and having in mind the condition (\ref{without-delta}), we find
\begin{equation}
(X-8\pi)[T^\gamma _\gamma] +3Z+\epsilon Q =0. \label{condition}
\end{equation}

On the other hand, and according to \cite{Szekeres}, the internal OS metric can be written
\begin{equation}
ds^2_{1}= -d\tau^2+\tau^{4/3}(d\rho^2+\rho^2d\Omega^2), \label{Einstein-De Sitter}
\end{equation}
that is, the well-known Einstein-de Sitter metric, while the external metric, the Schwarzschild one, in suitable $(T,R)$ coordinates \cite{Lemaitre-1933}, becomes
\begin{equation}
ds^2_{2}=-dT^2+\frac{4}{9} \frac{\rho_0^2}{(T+R)^{2/3}}dR^2+\rho_0^2(T+R)^{4/3}d\Omega^2, \label{Schwarzschild}
\end{equation}
with $\frac{2}{9}\rho_0^3=m$, $m$ being the total mass, and $R=0$ the boundary surface (see \cite{Szekeres}). That is, the boundary equation 
$\Phi=0$ reduces now to $R=0$. Following \cite{Szekeres} let us consider the coordinate transformation in the internal case:
\begin{equation}
\tau=T-\frac {2}{9} \frac{\rho_0^2}{T^{5/3}}R^2, \quad \rho=\rho_0 \Big(1+\frac {2}{3}\frac{R}{T}-\frac{1}{9}\frac{R^2}{T^2}\Big). \label{coordinate-trans}
\end{equation}

Then, after some calculation, it can be seen that the entire metric, the internal and the external, in the coordinates $(T,R)$, is as announced a 
${\cal C}^1$ class metric.

On the other hand, in the coordinates $(\tau,\rho)$, the only nonvanishing component of the energy-momentum tensor is obviously the (00) component, that is, the matter density $\mu(\tau)$. Then, from the coordinate transformation (\ref{coordinate-trans}), it is easy to see that the same holds for the coordinates $(T,R)$ on the boundary surface $R=0$. That is, in these coordinates,
\begin{equation}
T_{\alpha\beta}|_{R=0} = \delta_\alpha^0 \delta_\beta^0 T_{00}|_{R=0},
\label{Too}
\end{equation}
such that $T_{00}|_{R=0}=T^{00}|_{R=0}=\mu|_{R=0} \equiv \mu|_0$.  For the step at $R=0$, we have:
\begin{equation}
[T_{\alpha\beta}] = \delta_\alpha^0 \delta_\beta^0 [\mu]. \label{energy-momentum}
\end{equation}
Further, the unit vector $n_\alpha$ becomes in the present case
\begin{equation}
n_\alpha = \frac{(0,1,0,0)}{\sqrt{g^{\beta \gamma}|_0\partial_\beta R \partial_\gamma R}}=\sqrt{g_{11}|_0} \, (0,1,0,0). \label{n-vector}
\end{equation}
with  $g_{11}|_0$ the corresponding metric component of (\ref{Schwarzschild}) calculated  for $R=0$,  that is 
\begin{equation}\label{g11R0}
g_{11}|_0 = \frac{4}{9} \frac{\rho_0^2}{T^{2/3}}. 
\end{equation}
Eq. (\ref{n-vector}) means in particular that, since the vector $v_\alpha$ is by definition orthogonal to $n_\alpha$, it is necessarily $v_1=0$.

Then, let us consider the ten equations of (\ref{reducedEFE}). Having in mind (\ref{energy-momentum}), (\ref {n-vector}) and the fact that $v_1=0$, it is easy to verify that the only of these ten equations that do not reduce themselves to mere identities are the four diagonal ones $(\lambda,\mu=\lambda)$. These four equations, $\lambda = (0,1,2,3)$, become, respectively,
\begin{eqnarray}
X[\mu]-Z+Qv_0^2 & = & -8\pi [\mu], \label{first}\\ \nonumber\\
-X[\mu]+3Z+\epsilon Q & = & - 8\pi [\mu], \label{second}\\ \nonumber\\
Z g_{22}|_0+Qv_2^2b & = & - 8\pi [\mu] g_{22}|_0, \label{third}\\ \nonumber \\
Z g_{33}|_0+ Qv_3^2 & = & - 8\pi [\mu]g_{33}|_0, \label{fourth}
\end{eqnarray}
a particular solution being
\begin{equation}
X=-16\pi, \quad  Z=-8\pi [\mu], \quad  Q=0, \label{unique solution}
\end{equation}
that is, in accordance with (\ref{reducedEFE}), and having in mind that the functions $Y$ and $P$ of (\ref{general solution}) are arbitrary, we finally find for the corresponding $\kappa_{\lambda\mu}$:
\begin{equation}
\kappa_{\lambda\mu}= -16 \pi [S_{\lambda\mu}]+Y n_\lambda n_\mu +P(n_\lambda v_\mu +n_\mu v_\lambda ). \label{final kapa}
\end{equation}

We will see next that from these $\kappa_{\lambda\mu}$ values, in accordance with the relation (\ref{jumps-B}), we can obtain the corresponding jumps of the second derivatives of the OS metric in coordinates of Szekeres. Thus, since in this case the only non vanishing component of $n_\alpha$ is $n_1= \sqrt{g_{11}|_0}$, the only non vanishing jumps of these second derivatives will be the jumps of the $R$ second derivatives. That is,
\begin{equation}
\Big[\frac{\partial^2 g_{\alpha\beta}}{\partial R^2}\Big] = n_1^2 \, \kappa_{\alpha\beta}= g_{11}|_0 \, \kappa_{\alpha\beta}, \label{derivatives-kapa}
\end{equation}
that, in accordance with the metric (\ref{Schwarzschild}), becomes
\begin{equation}
\Big[\frac{\partial^2 g_{\alpha\beta}}{\partial R^2}\Big]  = \frac{4}{9} \frac{\rho_0^2}{T^{2/3}} \, \kappa_{\alpha\beta}, \label{derivatives-kapaB}
\end{equation}
or more explicitly (see (\ref{final kapa})),
\begin{equation}
\Big[\frac{\partial^2 g_{\alpha\beta}}{\partial R^2}\Big] = \frac{4}{9}\frac{\rho_0^2} {T^{2/3}} \Big(-16\pi[S_{\alpha\beta}]+
Yn_\alpha n_\beta+P(n_\alpha v_\beta+n_\beta v_\alpha)\Big).
 \label{derivatives-explicit}
\end{equation}

\vspace{2mm}

Then, by consistency, we must be able to choose $[\mu]$, $Y$, $P$, and $v_\alpha$ such that $v \cdot n=0$, $v^2=\epsilon$, in order to satisfy (\ref{derivatives-explicit}). But, according to Ref. \cite{Szekeres}, the internal OS metric in the $(T,R<0)$ coordinates, $g_{\alpha\beta}^{(1)}$, has the form
\begin{equation}
g_{\alpha\beta}^{(1)}=g_{\alpha\beta}^{(2)}+h_{\alpha\beta}, \label{metrics relation}
\end{equation}
where $g_{\alpha\beta}^{(2)}$ has the form of the external metric (see (\ref{Schwarzschild}), but now with $R<0$) and $h_{\alpha\beta}$ is a metric deformation such that $h_{\alpha\beta}$ and their first derivatives vanish at $R=0$ for all $T>0$. Because of this form of $g_{\alpha\beta}^{(1)}$, the jump $[\partial^2 g_{\alpha\beta}/\partial R^2]$ becomes
\begin{equation}
\Big[\frac{\partial^2 g_{\alpha\beta}}{\partial R^2}\Big]= - \frac{\partial^2 h_{\alpha\beta}}{\partial R^2}\Big|_{R=0}. \label{h-metric}
\end{equation}
In the Appendix, we give the explicit expressions of these $h_{\alpha\beta}$ components.

Thus, the equation that must be consistently satisfied is
\begin{eqnarray}
\frac{\partial^2 h_{\alpha\beta}}{\partial R^2}\Big|_{R=0} & = & -\frac{4}{9}\frac{\rho_0^2}{T^{2/3}}\Big (-16\pi[T_{\alpha\beta}]  
+ 8\pi [T^\gamma_\gamma] g_{\alpha\beta} \nonumber \\ & & + Y n_\alpha n_\beta +P(n_\alpha v_\beta+n_\beta v_\alpha)\Big). \label{consistent}
\end{eqnarray}
Then, let us see in detail that this is the case. To begin with, the three equations corresponding to the $(\alpha,\beta)$ values, $(0,2)$, $(0,3)$, and $(2,3)$ become identically equal to zero. Then, the equation $(0,1)$ gives $P v_0\ne 0$ on account that
\begin{equation}\label{zero-one}
\frac{\partial^2 h_{01}}{\partial R^2}\Big|_{R=0}  = P v_0 n_1= P v_0 \sqrt{g_{11}|_0},
\end{equation}
while the equations $(1,2)$, $(1,3)$, give  respectively $Pv_2=0$ and $Pv_3=0$. Hence, we obtain $v_2=v_3=0$. This means that $\epsilon = v^2 \equiv g^{\alpha \beta}v_\alpha v_\beta = g^{00}|_0 v_0^2=-v_0^2=-1$. It still remains the equations $(0,0)$, $(2,2)$, and $(3,3)$, that is, respectively,
\begin{eqnarray}
\frac{\partial^2 h_{00}}{\partial R^2}\Big|_{R=0} & = & 8\pi[\mu]  \, g_{11}|_0, \label{zero}\\
\frac{\partial^2 h_{22}}{\partial R^2}\Big|_{R=0} & = & 8\pi [\mu] \, g_{11}|_0 \, g_{22}|_0 , \label{two}\\
\frac{\partial^2 h_{33}}{\partial R^2}\Big|_{R=0} & = & 8\pi [\mu] \, g_{11}|_0 \, g_{33}|_0. \label{three}
\end{eqnarray}
It is easy to see that Eqs. (\ref{two}) and (\ref{three}) are equivalent. Then, we are left with Eqs. (\ref{zero}) and (\ref{two}). In order to make them compatible, we must have
\begin{equation}
\frac{\partial^2 h_{22}}{\partial R^2}\Big|_{R=0}  =  \frac{\partial^2 h_{00}}{\partial R^2}\Big|_{R=0} g_{22}|_0,   \label{compatible}
\end{equation}
which becomes true according to the values of  $\partial^2 h_{22}/\partial R^2|_0$ and $\partial^2 h_{00}/\partial R^2|_0$ (see Appendix. From (\ref{zero-one}) and (\ref{tercera}), taking into account (\ref{g11R0}), one obtains that $P v_0 = (8/9) \rho_0 T^{-7/3}$. In all, it remains Eq. (\ref{zero}) which 
gives $[\mu]=-\mu$, the minus matter density, as a function of the time $T$; then, from (\ref{g11R0}), (\ref{zero}) and (\ref{primera}) we have
\begin{equation}\label{densitat-EdeS}
\mu = \frac{1}{6 \pi T^2}, 
\end{equation}
as required by the Einstein-de Sitter solution (notice that, according to (\ref{coordinate-trans}), $T = \tau$ on the boundary surface $R=0$).

Finally, we still have to consider the $(1,1)$ component of Eq. (\ref{consistent}). This component will be satisfied  fitting the $T$ function $Y$.

The conclusion of all this is that the OS metric, in Szekeres coordinates, is a particular case of our assumed family of metrics satisfying the Lichnerowicz postulate (2). But, in order to actually conclude this, we need to arrive at three previous partial conclusions which do not appear in Ref. \cite{Szekeres}. The first of these three conclusions is that an atlas has to exist where the coordinates used, $(T, R)$, are admissible coordinates. Notice that the used polar Szekeres coordinates need at least two charts to form this atlas. These two charts could be, for instance, the ones tied to two given polar axes, respectively. But, in the region where these two charts intersect the Jacobian of the corresponding polar coordinates transformation is completely smooth, and so the coordinates considered, $(T, R)$, are admissible as we wanted to prove.

The second previous conclusion to achieve is that, in the $(T, R)$ coordinates, the first derivatives of the metric are ${\cal C}^2$ class. Finally, the third one is to show that the Jacobian matrix of the coordinate transformation $T=T(\tau, \rho)$, $R=R(\tau, \rho)$, does not vanish as it must be required. Both partial conclusions are reached in the Appendix, where two typographical mistakes in Ref. \cite{Szekeres} are also corrected.
%


\section{The case of the noncommutative second partial derivatives of the metric}
\label{sec:6}

In \cite{Kuntz}, an interesting generalization of general relativity is presented where the second partial derivatives of the metric do not commute in some regions of the total space-time: Here and for the sake of simplicity, in some boundary surface $\Phi=0$. That is, along this boundary, the Schwartz theorem does not necessarily holds. More specifically, the second partial derivatives of the metric exist everywhere but are not continuous along $\Phi=0$. Further, the metric is assumed to be ${\cal C}^1$ class everywhere.

In this theory, the definition of the Riemann curvature tensor, $R^{\mu}_{\ \nu\rho\sigma}$, remains the same, that is, for any smooth enough vector field, $v_{\nu}$, the noncommutativity of the covariant partial second derivatives writes down:
 \begin{equation}
 \nabla_{\sigma}\nabla_{\rho}v_\nu -\nabla_{\rho}\nabla_{\sigma}v_\nu =
  v_\mu R^{\mu}_{\ \nu\rho\sigma}. \label{curvature-def}
 \end{equation}

 From this definition, we obtain for the discontinuous part of the curvature tensor, $R^D_{\mu\nu\rho\sigma}$,
 \begin{equation}
R^D_{\mu\nu\rho\sigma} =\frac{1}{2}( \partial^2_{\sigma\mu}g_{\nu\rho} - \partial^2_{\sigma\nu}g_{\mu\rho}  -  \partial^2_{\rho\mu}g_{\nu\sigma} +
\partial^2_{\rho\nu}g_{\mu\sigma}) +
\frac{1}{2}\partial^2_{[\rho \sigma]}g_{\mu\nu}, \label{linear-curvature}
\end{equation}
where $[\cdot]$ denotes index antisymmetrization or,   more specifically, $\partial^2_{[\sigma \rho]} g_{\mu\nu} \equiv \frac{1}{2}(\partial_\sigma\partial_\rho - \partial_\rho \partial_\sigma) g_{\mu\nu}$.

But now the majority of the symmetries of $R_{\mu\nu\rho\sigma}$ are broken with the exception of the skew symmetry of the last two indices $R_{\mu\nu\rho\sigma}= -R_{\mu\nu\sigma\rho}$. This implies the existence of two different Ricci tensors, but only the one defined as
\begin{equation}
R_{\nu\sigma} \equiv g^{\mu\rho}R_{\mu\nu\rho\sigma}, \label{Ricci}
\end{equation}
has a nonvanishing associated contracting scalar $R \equiv g^{\rho\sigma}R_{\rho\sigma}$.

Nevertheless, we can still define the dual curvature tensor
\begin{equation}
\ast R_{\mu\nu\rho\sigma} \equiv \frac{1}{2}\eta_{\mu\nu}^{\ \ \alpha\beta} R_{\alpha\beta\rho\sigma}, \label{dual-curvature}
\end{equation}
where $\eta$ is the Levi-Civita tensor, and from it we can define the dual Ricci tensor
\begin{equation}
\tilde {R}_{\nu\sigma}\equiv g^{\mu\rho}(\ast R_{\mu\nu\rho\sigma}) , \label{dual-Ricci}
\end{equation}
and the dual Ricci scalar
\begin{equation}
\tilde {R}\equiv g^{\nu\sigma}\tilde {R}_{\nu\sigma} , \label{dual-scalar}
\end{equation}

From the two scalars, $R$ and $\tilde {R}$, plus the corresponding matter Lagrangian density, a suitable action is built in \cite{Kuntz}, whose standard variations lead to the following dynamical equations generalizing the Einstein field equations
\begin{eqnarray}
R_{(\mu\nu)}-\frac{1}{2}R g_{\mu\nu} & = &  \frac {1}{M_{p}^{2}}T_{\mu\nu}, \label{first-eq.}\\
R_{[\mu\nu]}+\frac{M^2}{M_{p}^{2}}\tilde {R }_{[\mu\nu]} & = &  \frac {1}{M_{p}^{2}}\nabla_{\alpha}S^{\alpha}_{ \ \ \mu\nu}, \label{second-eq.}
\end{eqnarray}
where $(\cdot)$ denotes index symmetrization, that is, $R_{(\mu\nu)} \equiv \frac{1}{2}(R_{\mu\nu}+ R_{\nu\mu})$,  and $[\cdot]$ has been defined above, $M_p$ is the Planck mass, and $M$ is a mass parameter to be determined by observations. Finally, $T_{\alpha\beta}$ is the standard energy-momentum tensor, that is related to the canonical one, $\Theta_{\alpha\beta}$,  by the usual assumption (cf. Eq. (42) in Ref. \cite{FoRo-2004}):
\begin{equation}\label{dosTes}
T^{\mu \nu} = \Theta^{\mu \nu} + \frac{1}{2} \nabla_\alpha (S^{\alpha \mu \nu} + S^{\mu \nu \alpha} - S^{\nu \alpha \mu})
\end{equation}
where $S^{\alpha \mu\nu} = - S^{\alpha \nu\mu}$ denotes the spin current.  In searching for the failure of parity symmetry in certain extensions of the general relativity theory, a gravitational action term, similar to the one considered in \cite{Kuntz}, has been introduced (see, for instance, 
Ref. \cite{HoMuSa-1980}) in the past. 

Notice that, out of the boundary surface, these generalized Einstein field equations become the standard EFEs, that is, (\ref{first-eq.}) with $R_{(\mu\nu)}$ identified with $R_{\mu\nu}$.

Now, it is interesting to remark that the discontinuous part of $R_{(\mu\nu)}$ that we denote here by $R^D_{(\mu\nu)}$ contains both, anticommutators $\partial^2_{(\alpha\beta)}g_{\gamma\delta}$ and commutators $\partial^2_{[\alpha\beta]}g_{\gamma\delta}$, while $R_{[\mu\nu]}$ only contains commutators. Furthermore, $\tilde {R}_{[\mu\nu]}$ only contains commutators too. More precisely, having in mind that $g^{\lambda\kappa}$ is symmetric in the indices $\lambda$, $\kappa$, we easily obtain:
\begin{eqnarray}
R^D_{(\mu\nu)} & = & \frac{1}{2}g^{\lambda\kappa}(\partial^2_{(\nu\lambda)}g_{\mu\kappa} + \partial^2_{(\mu\lambda)}g_{\nu\kappa} 
- \partial^2_{(\nu\mu)}g_{\lambda\kappa} - \partial^2_{(\lambda\kappa)}g_{\mu\nu}) \nonumber \\
& & + \frac{1}{2}g^{\lambda\kappa}\partial^2_{[\kappa\nu]}g_{\lambda\mu}+
\frac{1}{2}g^{\lambda\kappa}\partial^2_{[\kappa\mu]}g_{\lambda\nu}, \label{Rsymmetric}\\
R_{[\mu\nu]} & = & \frac{1}{2}g^{\lambda\kappa}
(\partial^2_{[\mu \nu]}g_{\lambda\kappa} + \partial^2_{[\nu \lambda]}g_{\mu\kappa}+
\partial^2_{[\kappa\mu]}g_{\lambda\nu}) \nonumber \\ 
& & + \frac{1}{2}g^{\lambda\kappa}\partial^2_{[\kappa\nu]}g_{\lambda\mu}-
\frac{1}{2}g^{\lambda\kappa}\partial^2_{[\kappa\mu]}g_{\lambda\nu} . \label{Rantisymmetric}
\end{eqnarray}

Finally, having in mind that the Levi-Civita tensor is completely antisymmetric, we easily obtain:
\begin{eqnarray}
\tilde{R}_{[\mu\nu]} & = & \frac{1}{8}\eta^{\sigma \ \alpha\beta}_{\ \mu} (\partial^2_{[\sigma\beta]}g_{\alpha\nu} - \partial^2_{[\sigma\alpha]}g_{\beta\nu}) \nonumber \\ 
& & - \frac{1}{8}\eta^{\sigma \ \alpha\beta}_{\ \nu} (\partial^2_{[\sigma\beta]}g_{\alpha\mu} - \partial^2_{[\sigma\alpha]}g_{\beta\mu}). \label{Rtilde-anti}
\end{eqnarray}

Equations (\ref{Rsymmetric})-(\ref{Rtilde-anti}) are neither obtained nor commented on in the cited Ref. \cite{Kuntz}, but they have the interest of showing some decoupling of the dynamical equations (\ref{first-eq.}) and (\ref{second-eq.}): The first of these equations involvs both the anticommutators and the commutators of the second partial derivatives of the metric, while the second one involves exclusively the corresponding commutators. But the structure of these two dynamical equations is very different: In the first one the nonlinear terms in the first partial derivatives of the metric appear, while in the second one only the second partial derivatives appear (linearly). However, if in some given physical situation, these commutators or anticommutators, or both, in the two sides of the boundary surface, had any physical significance, the same or even more could be said of the occasional finite jumps of the second partial derivatives of the metric. These jumps have been studied in the preceding sections in the frame of general relativity for ${\cal C}^1$ class metrics: The same class that has been assumed in the present section. For these jumps, the dynamical equations (\ref{first-eq.}) and (\ref{second-eq.}) would give
\begin{equation}
[R_{(\mu\nu)}]-\frac{1}{2}[R] g_{\mu\nu} =   \frac {1}{M_{p}^{2}} [T_{\mu\nu}], \label{first-jump}
\end{equation}
\begin{equation}
[R_{[\mu\nu]}]+\frac{M^2}{M_{p}^{2}}[\tilde {R }_{[\mu\nu]}] =  \frac {1}{M_{p}^{2}}[\nabla_{\alpha}S^{\alpha}_{\ \ \mu\nu}], \label{second-jump}
\end{equation}
where the big bracket $[\cdot]$ not involving indices stands for the corresponding jump.

The new second equation (\ref{second-jump}) does not introduce any simplification to the initial second equation (\ref{second-eq.}), but contrarily, the new first equation (\ref{first-jump}) is appreciably simpler than the initial equation (\ref{first-eq.}) since the nonlinear terms containing first partial derivatives of the metric have disappeared. In all, the new dynamical equations generalizing Einstein field equations, when reduced to the corresponding equations for the jumps of the second partial metric derivatives, that is to say, when reduced to Eqs. (\ref{first-jump}) and (\ref{second-jump}), become dramatically simplified: The first partial metric derivatives disappear everywhere, and, as a consequence, these equations become linear in the remaining partial metric derivatives, the second ones. This could be relevant when trying to compare the new theory to the observation for strong gravitational fields  beyond the particular linearized case.

Before ending the paper, let us come back to the Hadamard discontinuities in Sec. \ref{sec:4}  and more specifically to Eq. (\ref{jumps}). In the case of the present section, this equation remains true but we cannot anymore conclude the following Eq. (\ref{jumps-B}) anymore, since now the jumps $[\partial^2_{\lambda\mu}g_{\nu\kappa}]$ are no longer symmetric in the indices $\lambda$ and $\mu$.


\section{Summary of findings}
\label{sec:7}

In this section the main findings of the paper are summarized. Two different topics have been considered. 
The first one is approached in Secs. \ref{sec:2}--\ref{sec:4}, and the second one in Sec. \ref{sec:6}, while Sec. \ref{sec:5} 
is devoted to presenting a notorious example of the first topic.

This topic deals with the solutions of the EFEs whose energy-momentum tensor has a bounded support, where the tensor is everywhere continuous except for a finite step across the boundary surface, $\Phi = 0$, of the support. Because of this step, the second partial derivatives of these metric solutions show necessarily a finite jump across the boundary. Then, we have considered the family of these solutions whose metrics are continuous everywhere. The subfamily of this family of metrics, whose first partial derivatives are also continuous everywhere, is certainly nonempty, since the Newtonian potential solution of the Poisson equation for a finite bounded source is well known to be 
${\cal C}^1$ class. Even more, another metric of this subfamily is also presented in Sec. \ref{sec:5}. Hence, in Secs. \ref{sec:2}-\ref{sec:5} we have considered the set of this nonempty subfamily of metrics, that is, the set of all the above solutions of the EFE assumed to be  ${\cal C}^1$ class, and we have explored the necessary condition for the existence of such a kind of metrics. More specifically, using a well-known theorem from Hadamard \cite{Hadamard}, we have derived the necessary conditions that the jumps of the second partial derivatives of the metric, across the boundary surface, must satisfy in order to guarantee the ${\cal C}^1$ class character of the corresponding metrics. These necessary conditions are the Eqs. (\ref{EFE-kappa}) whose different actual solutions have been considered in the paper.

It is to be remarked that the stated necessary condition does not refer properly to Lichnerowicz's postulate (2), introduced just at the beginning of the Introduction, since this postulate, aside the ${\cal C}^1$ class character of the metric requires the use of admissible coordinates and also the ${\cal C}^2$ class character of the first derivatives of the metric. Nevertheless, in Sec. \ref{sec:5} (see the Appendix also) we have revisited the particular case of the OS metric referred to a global Szekeres coordinate system \cite{Szekeres}, where the metric shows its everywhere  ${\cal C}^1$ class character. Consequently, in this particular case, our necessary condition for it, Eq. (\ref{EFE-kappa}), has to be satisfied. Then, we have verified this condition by building for the case the corresponding solution of Eq. (\ref{EFE-kappa}). Further, we have proved that these coordinates are admissible ones (Sec. \ref{sec:5}), and also the ${\cal C}^2$ class character of the first derivatives of the metric (Appendix),
thus, proving that this particular case is an example of a metric satisfying the entire Lichnerowicz`s postulate (2). Also, we have taken advantage of this renewed visit to the OS metric to prove that the Jacobian of the coordinate transformation, leading from the original Gaussian coordinates of this metric to the Szekeres ones, never vanishes as due. Finally, we have given the explicit expression of the OS metric in the new coordinates, and we have pointed out two minor missprints in the referred paper \cite{Szekeres}.

The last findings of the present paper are contained in Sec. \ref{sec:6}. In this section, we have considered a recent generalization of the EFE \cite{Kuntz} to a case where the mixed second partial derivatives of the metric, at the boundary surface, $\Phi = 0$, exist but are not symmetric, i.e., the Schwartz theorem is no longer valid. In other words, the existing second derivatives of the metric are not continuous at least at one of the two sides of $\Phi = 0$. In this generalization of the EFE, the metric is assumed to be ${\cal C}^1$ class.

The new field equations that generalize the EFE are (\ref{first-eq.}) and (\ref{second-eq.}) in Sec. \ref{sec:6}. Then, in this section, we have shown that Eqs. (\ref{first-eq.}) and (\ref{second-eq.}) decouple in the following sense: While  Eq. (\ref{first-eq.}) depends on both, the commutators and the anticommutators of the second derivatives of the metric, Eq. (\ref{second-eq.}) depends only on the commutators. This can be seen from Eqs. (\ref{Rantisymmetric}), (\ref{Rtilde-anti}), giving the explicit expressions for $R_{[\mu\nu]}$ and $\tilde{R}_{[\mu\nu]}$, respectively. Further, we have pointed out that the structures of Eqs. (\ref{first-eq.}) and (\ref{second-eq.}) are very different: While Eqs. (\ref{first-eq.}) include the presence of the first derivatives of the metric, Eqs.  (\ref{second-eq.}) do not. So Eqs. (\ref{second-eq.})  are simpler than Eqs. (\ref{first-eq.}). Then, following formally what we have done in the sections before \ref{sec:6}, in this section we have focused our attention, not on the generalized field equations, but on their jumps, that is, on Eqs. (\ref{first-jump}) and (\ref{second-jump}), which become algebraic equations for the jumps of the above metric commutators and anticommutators. These equations are defined across the boundary surface $\Phi = 0$ and become particularly simple since they do not depend on the first derivatives of the metric: Eqs. (\ref{second-jump}), because preserve this nondependence property already present in its antecedent, Eq. (\ref{second-eq.}); Eq. (\ref{first-jump}), like a gain of simplicity tied to their jumping character. Notice that the reduction of the original generalized field equations to their jumping counterpart is not a mere artifact to accede to simpler field equations, since generally speaking the jumps of the second derivatives of the metric, by themselves, could become particularly interesting.

%

\begin{acknowledgments}
This work has been partially supported by the Spanish ``Ministerio de Econom\'{\i}a y Competitividad'' and the ``Fondo Europeo de Desarrollo Regional'' MINECO-FEDER. It was initiated during the Project No. FIS2015-64552-P.
\end{acknowledgments}


\appendix

\section{The Oppenheimer-Snyder metric in coordinates of Szekeres}
\label{ap-A}

In this appendix we summarize the intermediate steps of the Ref. \cite{Szekeres} computations allowing us to obtain the announced compatibility relation (\ref{compatible}). In relation to this reference, we also take advantage of the occasion to point out two misprints, and also to verify, as due, that the Jacobian of the coordinate transformation going from $(T, R)$ coordinates to the Gaussian ones $(\tau, \rho)$ never vanishes, and finally to show that the first derivatives of the OS metric in $(T, R)$ coordinates are piecewise ${\cal C}^2$  class.

In Ref. \cite{Szekeres}, Szekeres obtained a coordinate system $(T, R)$ in which the OS metric exhibits its ${\cal C}^1$ class character everywhere and, in particular, through the junction surface, $R=0$. The OS metric represents a homogeneous cloud of dust matter radially collapsing. The Jacobian determinant, $J(T,R)$, of the coordinate transformation (\ref{coordinate-trans}), from 
$(T, R)$ coordinates to comoving 
Gaussian coordinates $(\tau, \rho)$, is given by
\begin{equation}\label{jacobia}
J(T, R)= \frac{2}{3} \rho_0 \frac{1}{T} (1- \frac{1}{3}\frac{R}{T})(1- \frac{2}{27} \rho_0^2 \frac{R^2}{T^{8/3}}), 
\end{equation}
which does not vanish on the space-time region
\begin{equation}\label{domini}
-3  (\sqrt{2} -1) T \leq R \leq 0, \quad 3 T^{4/3} > - \sqrt{2} \rho_0 R,
\end{equation}
defining the whole coordinate domain. For the sake of conciseness, angular  coordinates $\theta$ and $\phi$ adapted to the spherical symmetry are omitted. Using  $(T, R)$ coordinates, the interior Einstein--de Sitter metric (\ref{Einstein-De Sitter}), $ds_1^2$, of the collapsing OS scenario, can be written as a ${\cal C}^1$ class metric deformation of the Schwarzschild metric (\ref{Schwarzschild}), $ds_2^2$, expressed in terms of Lema{\^{\i}}tre coordinates. These coordinates are  adapted to a congruence of free-falling radial observers, which are asymptotically at rest at spatial infinity. Using the notation 
\begin{equation} \label{ds1ds2}
ds_1^2 = ds_2^2 + \rho_0^2 \, d\sigma^2,
\end{equation}
the line element deformation $d\sigma^2$ providing the inside OS solution, in the referred $(T, R)$ coordinates,  takes the expression
\begin{equation} \label{deformacio}
d\sigma^2 = R^2 T^{-8/3} A dT^2 + B dR^2 + 2 R T^{- 5/3} C dT dR + D d \Omega^2, 
\end{equation}
where $A, B, C$ and $D$, are functions of $T$ and $R$ given by
\begin{eqnarray}
A(T,R) & \equiv & -\frac{2}{9}\Big[\frac{10}{3}(1+\frac{25}{9} F^2)- 2(1-F^2)^{4/3} H^2\Big], \label{A-function}\quad \\
B(T,R) & \equiv & -\frac{4}{9}\Big\{T^{-2/3} \Big[ 2 F^2 -  (1- F^2)^{4/3} H^2 \Big] \nonumber \\ & & + (T+R)^{-2/3} \Big\}, \label{B-function}\\
C(T,R) & \equiv & \frac{4}{9} \Big[1 + \frac{5}{3} F^2 - (1- F^2)^{4/3} H^2 \Big], \label{C-function}\\
D(T,R) & \equiv & T^{4/3} (1- F^2)^{4/3} \Big(1 + \frac{2}{3} \frac{R}{T} - \frac{1}{9}\frac{R^2}{T^2}\Big)^2 \nonumber \\ & & -   (T + R)^{4/3}, \label{D-function}
\end{eqnarray}
with 
\begin{equation}\label{FH-functions}
F^2 \equiv \frac{2}{9} \rho_0^2 R^2 T^{-8/3}, \quad H^2 \equiv \Big(1 - \frac{1}{3} \frac{R}{T}\Big)^2.
\end{equation}

Now, as stated, let us show that the first derivatives of the OS metric in $(T, R)$ coordinates are piecewise ${\cal C}^2$ class. In other words, out of the boundary surface $R=0$, that is to say, for $R \neq 0$, we must show that the corresponding second and third derivatives of this metric are continuous. More specifically, we must show this continuity both for $R > 0$ and for $R < 0$: In the first region, $R > 0$, the corresponding metric is the outer Schwarzschild metric in the $(T, R)$ coordinates, i.e., the metric (\ref{Schwarzschild}); in the second region, $R < 0$, the metric 
(\ref{deformacio})-(\ref{FH-functions}). In both regions an apparent discontinuity of these second and third derivatives at $T + R = 0$  is present, but this coordinate condition describes the collapsing singularity that does not belong to the present differentiable manifold. We can directly see that there are no more apparent or actual discontinuities of the second and third derivatives of the metric (\ref{Schwarzschild}). We are then left with the region $R < 0$. The simple examination of Eqs. (\ref{A-function})-(\ref{FH-functions}) seems to show two possible cases of discontinuity in those derivatives and only two cases: for $T = 0$ and $F^2 = 1$, respectively.

Let us first try the case $T = 0$. Let us consider then the defining coordinate domain (\ref{domini}). More specifically, the second of these inequalities, that is, the strict inequality $3 T^{4/3} >  - \sqrt{2} \rho_0 R$. Since we are now in the region $R < 0$, this strict inequality says that this time value, 
$T = 0$, is unattainable. Now, let us go to the case $F^2=1$. Because of the $F^2$ definition in the first equation of (\ref{FH-functions}), $F^2 = 1$ becomes $3 T^{4/3} = - \sqrt{2} \rho_0 R$, that again becomes unattainable because of the above strict inequality.

In all, out of $R = 0$, the second and third derivatives of the OS metric in $(T, R)$ coordinates are continuous, as we wanted to prove.

Then, before ending this Appendix, let us come back to the pending compatibility relation (\ref{compatible}). From the metric (\ref{deformacio}), and using the latter definitions, the second radial derivatives of the nonvanishing $h_{\mu\nu}$ components of the metric deformation $\rho_0^2 \, d\sigma^2$ can be straightaway evaluated on  the junction boundary $R=0$,  giving the result:
\begin{eqnarray}\label{parcial-segon}
\frac{\partial^2 h_{00}}{\partial R^2}\Big|_{R=0} & = & - \frac{16}{27} \rho_0^2 \, T^{-8/3}, \label{primera}\\
\frac{\partial^2 h_{11}}{\partial R^2}\Big|_{R=0} & = &  - \frac{32}{81} \rho_0^2 \, T^{-8/3}(1 + \frac{5}{3} \rho_0^2 \, T^{-2/3}), \label{segona} \\
\frac{\partial^2 h_{01}}{\partial R^2}\Big|_{R=0} & = & \frac{16}{27} \rho_0^2 \, T^{-8/3}, \label{tercera}\\
\frac{\partial^2 h_{22}}{\partial R^2}\Big|_{R=0} & = & - \frac{16}{27} \rho_0^4 \, T^{-4/3}\label{quarta},
\end{eqnarray}
and $\frac{\partial^2 h_{33}}{\partial R^2}\Big|_{R=0} =  \sin^2\theta \,  \frac{\partial^2 h_{22}}{\partial R^2}\Big|_{R=0}$, indeed, where  $T, R, \theta, \phi$ coordinates are referred by the indices $0, 1, 2, 3$, respectively. Then, the required compatibility relation  (\ref{compatible}) is identically satisfied by virtue of (\ref{primera}) and (\ref{quarta}).

Finally, the first of the two mentioned misprints in Ref. \cite{Szekeres} refers to the Eq. (6) of this reference, in whose left-hand side we must put $\rho$ instead of $r$. The second misprint refers to the equation giving $ds^2$ in the page 189 of \cite{Szekeres}, where we must change $R < 0$ by $R > 0$, and reciprocally.


%

\end{document}